\documentclass{aa}
\usepackage{psfig}
\usepackage{graphicx}
\usepackage{txfonts}

\begin{document}

  \title{The soft X-ray Cluster-AGN spatial cross-correlation function in the 
                        {\em ROSAT}-NEP survey.}

    \author{N.~Cappelluti\inst{1}\and H. B\"ohringer\inst{1}\and P. Schuecker\inst{1}
      \and E. Pierpaoli\inst{2,6}\and C.~R.~Mullis\inst{3}\and I.~M.~Gioia\inst{4}
      \and J.~P.~Henry\inst{5,1}
        }

%\authorrunning{N. Cappelluti et~al.}
%\titlerunning{Soft X-ray Cluster-AGN spatial cross-correlation function}

  \institute {Max Planck Institute f\"ur Extraterrestrische Physik, Postfach 1312, 
               85741, Garching, Germany
    \and  California Institute of Technology, Mail Code 130-33, 
                     Pasadena, CA 91125, USA
    \and Department of Astronomy, University of Michigan, 918 Dennison, 
                    500 Church Street, Ann Arbor,   48109-1042	
    \and Istituto di Radioastronomia INAF, Via P. Gobetti 101, 40129 
                  Bologna, Italy
    \and Institute for Astronomy, University of Hawai'i, 2680 Woodlawn 
                 Drive, Honolulu, HI 96822
           \and   Physics and Astronomy Department, University of Southern California, 
	   Los Angeles, CA 90089-0484, USA   }

   \offprints{N. Cappelluti \email{cappelluti@mpe.mpg.de} }
   \date{Received  / Accepted  }

\abstract{X-ray surveys facilitate investigations of the environment
of AGNs. Deep {\textit Chandra} observations revealed that the AGNs
source surface density rises near clusters of galaxies.  The natural
extension of these works is the measurement of spatial clustering of
AGNs around clusters and the investigation of relative biasing between
active galactic nuclei and galaxies near clusters.}
{The major aims of this work are to obtain a measurement of the
correlation length of AGNs around clusters and a measure of the
averaged clustering properties of a complete sample of AGNs in dense
environments.}
{We present the first measurement of the soft X-ray cluster-AGN
cross-correlation function in redshift space using the data of the
{\em ROSAT}-NEP survey.  The survey covers $9\times9$ deg$^{2}$ around
the North Ecliptic Pole where 442 X-ray sources were detected and
almost completely spectroscopically identified.}
 {We detected a $>$3$\sigma$ significant clustering signal on scales
$s$ $\leq$50 $h_{70}^{-1}$ Mpc. We performed a classical
maximum-likelihood power-law fit to the data and obtained a
correlation length $s_{0}$=8.7$^{+1.2}_{-0.3}$ $h_{70}^{-1}$ Mpc and a
slope $\gamma$=1.7$^{+0.2}_{-0.7}$ ( 1$\sigma$ errors).}
{This is a strong evidence that AGNs are good tracers of the large
scale structure of the Universe.  Our data were compared to the
results obtained by cross-correlating X-ray clusters and galaxies. We
observe, with a large uncertainty, a similar behaviour of the AGNs
clustering around clusters similar to  the clustering of galaxies
around clusters.  }


 \maketitle

\section{Introduction}

 The current paradigm of galaxy formation assumes that all types of
galaxies reside in dark matter (DM) haloes, and that the properties of
these haloes determine to some extent the properties of the galaxies
inside (White \& Rees 1978). In general, the clustering amplitudes of
haloes depend on halo mass. The relation between the clustering
properties of both DM haloes and galaxies (biasing) should thus tell
us something about the physical processes leading to the formation and
evolution of galaxies. Clusters of galaxies are the highest peaks in
the global mass distribution of the Universe and should follow a
direct and  simple biasing scheme -- mainly related to the
underlying primordial Gaussian random field (Kaiser 1987). A simple
means towards a better understanding of galaxy biasing is thus
provided by studies of the relative biasing between galaxies and
clusters of galaxies.

  As a first step, previous investigations estimated two-point
statistics like the {\it auto}-correlation function (Mullis et al. 2004,
Gilli et al. 2005, Basilakos et al. 2005, Yang et al. 2006). They
could show that, in fact, AGNs trace the underlying cosmic large-scale
structure. In addition, the large-scale structure of X-ray selected
galaxy clusters could be studied in some detail (e.g. Schuecker et
al. 2001), but without investigating the link between clusters and
AGNs.

In the present paper, we concentrate on the study of the relative
clustering between X-ray selected AGNs and galaxy clusters. Our work
improves on most previous work on the large-scale structure of X-ray
selected AGNs in two important aspects. First, with the exception of
Mullis et al. (2004), our sample is the only one that is
spectroscopically complete (99.6\%). Gilli et al. (2005) used the CDFS
(~35\%) and the CDFN (50\%). The Basilakos et al. (2005) sample had almost no
spectroscopic redshifts. Yang et al. (2006) used the CLASXS sample
(52\% complete) and the CDFN (56\% complete). 
%Second, with the exception
%of Mullis et al. (2004) and part of Yang et al. (2006), we measure a
%three dimensional redshift space correlation function as opposed to
%deprojecting the two dimensional angular correlation function.
 
 Another motivation for our work is that over the last several years,
  X-ray observations revealed that a
 significant fraction of high-$z$ clusters of galaxies show
 overdensities of AGNs in their outskirts (i.e. between 3 $h_{70}^{-1}$
Mpc and 7 $h_{70}^{-1}$ Mpc from the center of the cluster) (Henry
et~al., 1991; Cappi et~al., 2001; Ruderman \& Ebeling 2005, Cappelluti
et~al., 2005, and references therein). These overdensities were
however detected in randomly selected archive targeted observations of
galaxy clusters. While these overdensities are highly significant (up
to 8$\sigma$) when compared to cluster-free fields, the incompleteness
of the samples does not allow drawing any conclusion about the average
clustering properties of AGNs around clusters. The majority of the
sources making these overdensities have no spectroscopical
identification and therefore any information on their spatial
clustering is lost.  More recently Branchesi et al. (2007) showed that
at high-$z$ the source surface density of AGNs significantly increases
even in the central regions of the clusters.  These results imply that
further progress will come from studying the three dimensional spatial
distribution of AGNs around clusters.  A natural way to characterize
this specific type of clustering is given by the three-dimensional
{\it cross}-correlation of AGNs and galaxy clusters, the computation
of which needs complete redshift information for all objects, which is
rare in X-ray surveys.

In this respect, the {\em ROSAT} North Ecliptic Pole (NEP) survey
(Henry et al., 2001,2006; Voges et al., 2001) is one of the few X-ray surveys covering a
sufficiently large volume with an almost complete follow-up
identification of AGNs and clusters (i.e. 440 sources
spectroscopically identified of 442 detected). This survey thus
provides a very useful basis for more precise investigations of the
relative clustering properties of these two types of objects.  

We organized the present paper in the following way. In Sect.\,2, we
describe the ROSAT-NEP survey data which we use for our investigations
of the spatial distribution of X-ray selected AGNs and galaxy
clusters. For the statistical analysis we estimate their
cross-correlation. A useful estimator for this statistic and the mock
samples needed for its determination are described in Sects.\,3 and 4,
respectively. The results are presented in Sect.\,5, and are discussed
in Sect.\,6. In this paper, we assume a (concordance)
Friedmann-Lemaitre Universe characterized by the Hubble constant given
in units of $h_{70}=H_0/(70\,{\rm km}\,{\rm s}^{-1}\,{\rm Mpc}^{-1})$,
the normalized cosmic matter density $\Omega_{\rm m}=0.3$, and the
normalized cosmological constant $\Omega_\Lambda=0.7$. Unless 
otherwise stated, errors are reported at the $1\sigma$ confidence level.
\section{The data}
The {\em ROSAT} NEP survey covers a region of $9\times9$ deg$^{2}$
around the North Ecliptic Pole (17$^{h}15^{m} < \alpha <
18^{h}45^{m}$, 62$^{o} < \delta < 71^{o}$) observed with the PSPC
proportional counter as part of the {\em ROSAT} All Sky Survey (Henry
et~al., 2001; Voges et~al., 2001) with a flux limit of
2$\times10^{-14} $erg cm$^{-2}$ s$^{-1}$ in the 0.5--2 keV energy
band. 442 X-ray sources were detected and 440 optically identified.
Spectroscopic redshift information is available for 219 AGNs and 62
clusters of galaxies. The clusters have redshifts $z\le$0.81 with a
median of 0.18 and the AGNs have $z\le$3.889 with a median of 0.4
(Fig. \ref{fig:dndz}). For the purpose of this work we selected all
the clusters and the 185 AGNs with $z\le$1 (Gioia et al. 2003). \\
Such a dataset was used also by Mullis et~al.~(2004) for the
calculation of the 3-D auto correlation function of X-ray selected
AGNs.  Mullis and collaborators find significant clustering on scales
smaller than $\sim$ 43 $h_{70}^{-1}$ Mpc with a correlation length of
$\sim$ 10.4 $h_{70}^{-1}$ Mpc, and a slope of the correlation best-fit
power law of $\gamma$=1.8.
\section{Cluster-AGN spatial cross-correlation}
 The cross-correlation function $\xi_{CA}$ of clusters and AGNs is
defined by the joint probability to find, at a distance $r$, one
cluster in the infinitesimal comoving volume element $\delta V_{C}$
and one AGN in the comoving volume element $\delta V_{A}$,
\begin{equation}
\delta P=n_{C}\,n_{A}\,[1+\xi_{CA}(r)]\,\delta V_{C}\,\delta V_{A},
\label{def}
\end{equation} 
\noindent
where $n_{C}$ and $n_{A}$ are the mean comoving number densities of
clusters and AGNs, respectively.  In calculating the differential
cross-correlation in redshift space we used an adapted version of the
Landy--Szalay estimator (Landy \& Szalay, 1993; see also e.g. Blake et
al., 2006),
\begin{equation}
\xi_{CA}(s)= \frac{D_{C}D_{A}-R_{C}D_{A}-R_{A}D_{C}+R_{A}R_{C}}{R_{A}R_{C}},
\label{eq:ccf}
\end{equation} 
where $D_{C}D_{A}$, $R_{C}D_{A}$, $R_{A}D_{C}$ and $ R_{A}R_{C} $ are
the normalized number of pairs  in the $i$-th redshift space
separation $s$ bin for the clusters data-AGNs data, cluster
random-AGNs data, AGNs random-clusters data and clusters random-AGNs
random samples, respectively. Using the symbols D and R to represent
the data and random samples, respectively, and C and A to identify
clusters and AGNs, respectively, the normalized pairs are expressed by
\begin{eqnarray*}
D_{C}D_{A}=\frac{n_{pair, D_{A}D_{C}}(s_{i})}{(N_{D_{A}}\times{N_{D_{C}}})},\,\,\,
R_{C}D_{A}=\frac{n_{pair, R_{C}D_{A}}(s_{i})}{(N_{R_{C}}\times{N_{D_{A}}})},\\
R_{A}D_{C}=\frac{n_{pair, R_{A}D_{C}}(s_{i})}{(N_{R_{A}}\times{N_{D_{C}}})},\,\,\,
R_{A}R_{C}=\frac{n_{pair, R_{A}R_{C}}(s_{i})}{(N_{R_{A}}\times{N_{R_{C}}})}.
\label{pairs}
\end{eqnarray*}
Here, $N_{D_{A}},N_{D_{C}},N_{R_{A}},N_{R_{C}}$ are the total numbers
of AGNs and clusters in the data and in the randomly generated
samples, respectively.  The quantities $n_{pair}$ represent,
adopting the symbolism used above, the actual number of pairs measured
in the random and data samples as a function of the redshift space
separation $s_{i}$.  The distances were computed assuming for the
position of the clusters the centroid of the X-ray emission while for
AGNs the optical positions were adopted.  In order to have a good
signal-to-noise ratio (SNR) the data were grouped in logarithmic bins
of $\Delta\log(s h_{70}^{-1}$~Mpc$)$=0.15.

\section{Random Samples}
\begin{figure}[!b]
\psfig{file=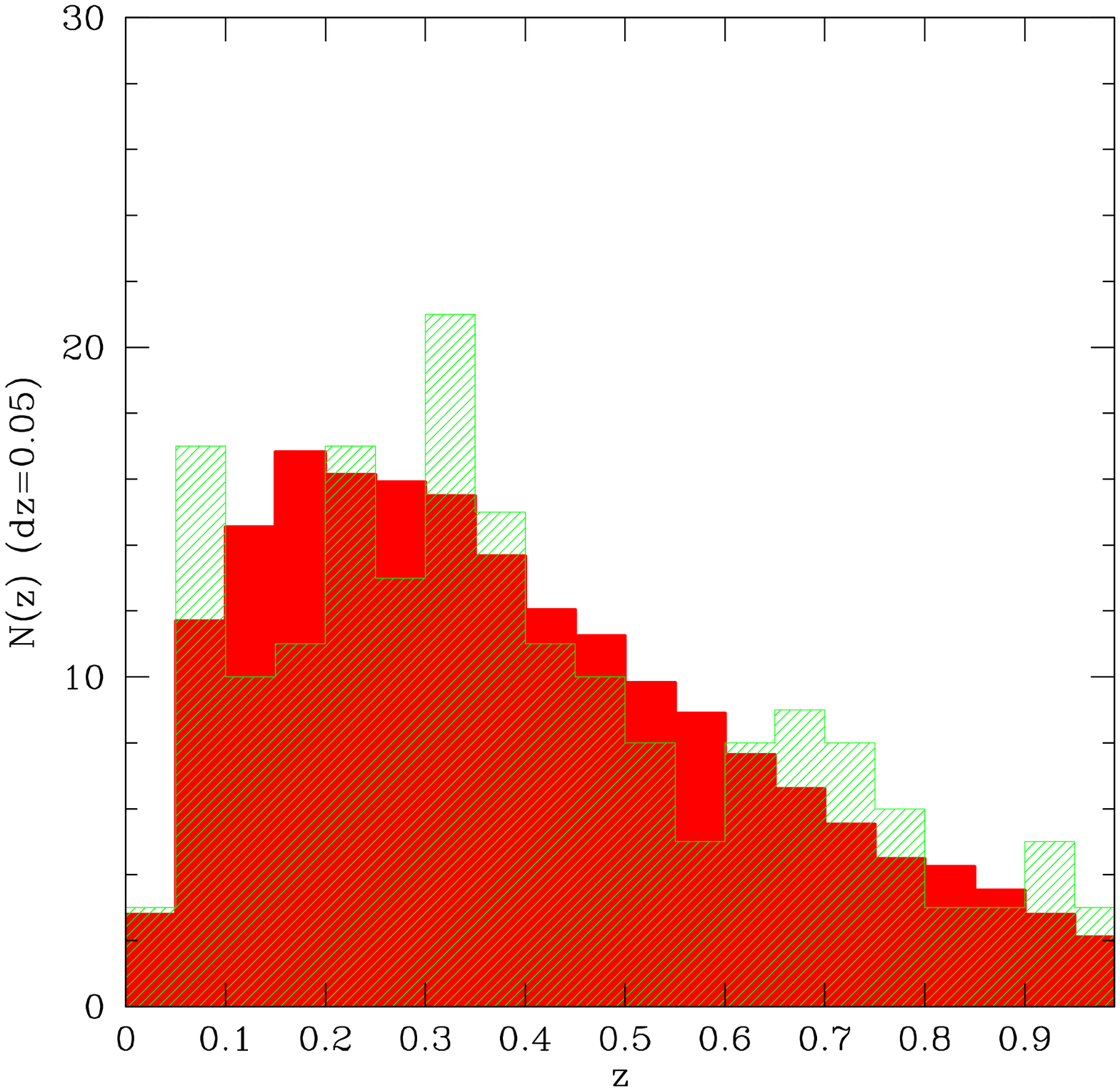,width=8.cm,height=6.5cm,angle=0}
\psfig{file=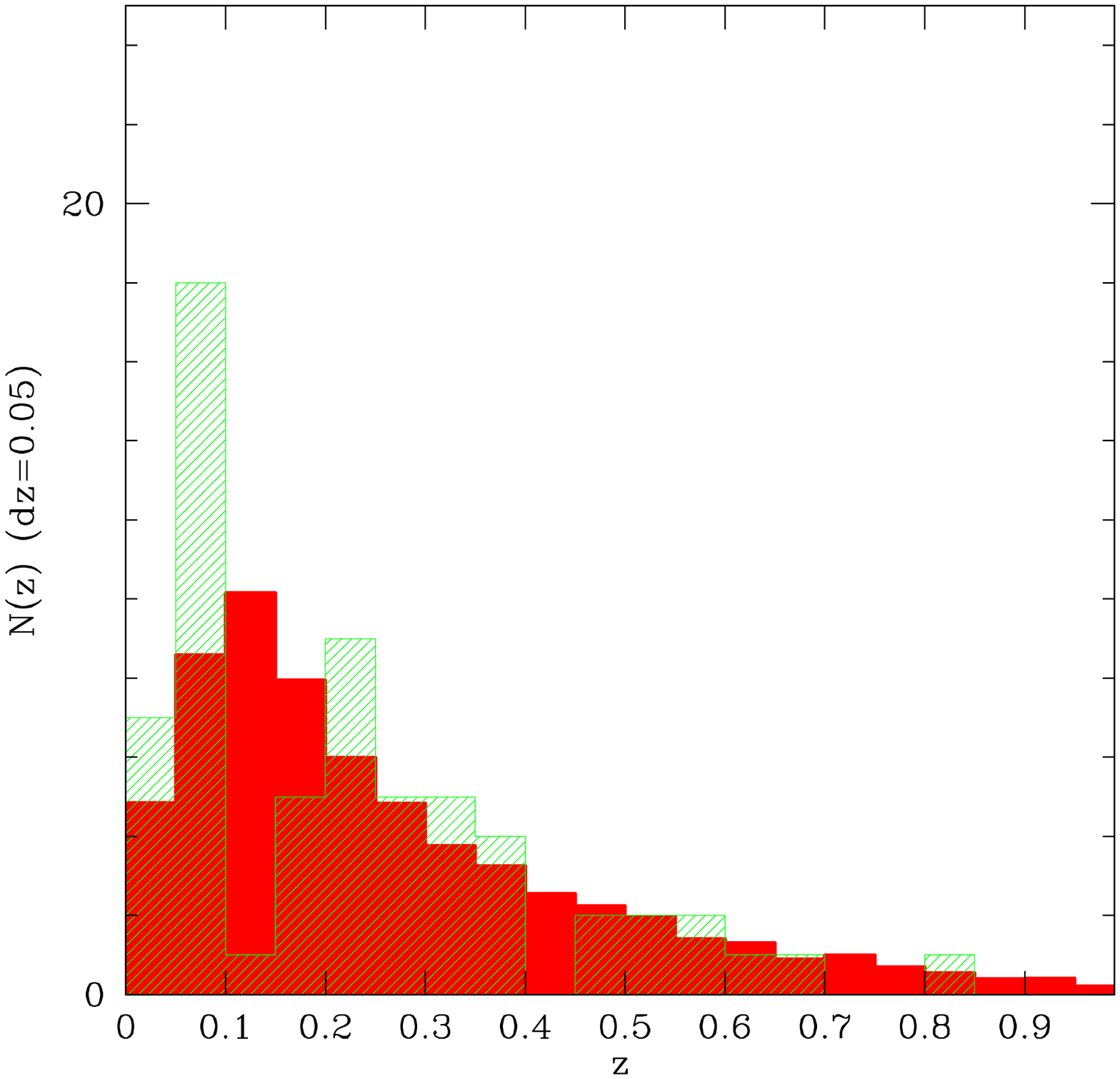,width=8cm,height=6.5cm,angle=0}
\caption{$Top~Panel$: The redshift distribution of the AGNs in the NEP 
survey  ($shaded~histogram$) and of the  randomly generated AGNs with the same selection 
effects  ($filled~histogram$). 
$Bottom~Panel$:  The redshift distribution of  the galaxy clusters in 
the NEP survey  ($shaded~histogram$) and of   randomly generated clusters ($filled~histogram$).
  The deviations at low $z$ between 
the data and the random sample are mainly caused by  the NEP supercluster of galaxies (Mullis et~al., 2001).}
\label{fig:dndz}
\end{figure}
Equation \ref{eq:ccf}  indicates that an accurate estimate of the 
distribution function of the random  samples is crucial in order to 
obtain a reliable  estimate of $\xi_{CA}$. Several effects must be taken 
 into account when generating a sample of objects in a flux limited survey.  
Simulated AGNs were randomly placed within the {\em ROSAT} NEP survey area.   
In order to reproduce the flux distribution of the real sample, we followed 
the method of Mullis et~al.~(2004).  In practice since the cumulative AGNs 
logN-logS source counts distribution can be  described by a power law, $S=kS^{-\alpha}$, with 
$\alpha=1.3$,  the differential probability scales as $S^{-(\alpha+1)}$. 
Using a transformation method (e.g. Press et al. 1986, chapter 7) we see that the 
random flux above  a certain X-ray flux $S_{lim}$ is distributed as 
$S=S_{lim}(1-p)^{\frac{1}{\alpha}}$, where $p$ is a random number uniformly 
distributed between 0 and 1 and $S_{lim}$=2$\times10^{-14}$ erg cm$^{-2}$ 
s$^{-1}$, i.e.  the flux limit of the NEP  survey. All  random AGNs 
with a flux lower than the flux limit map (see Fig. 4 in Henry et~al., 2006) at 
the source position were  excluded.
  In order to assign a redshift to these "sources" we computed  the predicted redshift 
distribution at the position of each accepted source. Once the flux limit 
 at the position where the source was randomly placed  is known, 
and   denoting with $\phi(L,z)$ the luminosity function, 
then the number of sources per redshift interval $dz$ is given by
\begin{equation}
N(z)dz=\int_{L_{min}}^{\infty}\int_{z}^{z+dz}\phi(L,z)\,dV(z)\,dL\,dz,
\end{equation}
where $L_{min}$ is the minimum luminosity  observable  at redshift $z$ 
with a  local flux limit S$_{lim}$ and $dV(z)$   the differential comoving volume 
element. The k-correction  does not play any role in the calculations 
  since we  assumed an average  spectral index $\Gamma$=2 
(as in Mullis et~al.~2004) for all AGNs. For the luminosity function 
$\phi(L,z)$  we took the luminosity-dependent density evolution 
(LDDE) best fit model of Hasinger et~al. (2005). 
A redshift was  then randomly assigned  to each source via Monte Carlo 
integration of the predicted redshift distribution. For  galaxy clusters 
we applied the same procedure assuming  a slope of the logN-logS distribution of 
  $\alpha$=1.3, as luminosity function an A-B evolving Schechter model (\cite{ros02}) 
 with the parameters obtained by \cite{mulxlf} and using a  sensitivity map specific for NEP
 clusters (\cite{hen06}).  
 Since for clusters of galaxies the k-correction is not negligible, we assigned to  the random 
 clusters an intrinsic spectrum according to a MEKAL spectral model with a 
plasma  temperature $kT$=3~keV and a metallicity $Z=0.3Z_{\odot}$.  
We  also applied  to L$_{min}$ a ``size-correction'' according to the results
of Henry et~al. (2006) in order to compensate for the missing flux in the X-ray photometry 
aperture and the variation of the angular dimensions of the object with $z$. 
Such a procedure was repeated until we populated the survey volume with 37200 
random AGNs and 12600 random clusters (i.e. 200 times more objects 
than in the real  data sample). The redshift distribution of the random 
cluster and AGN samples are plotted in Fig. \ref{fig:dndz} together 
with the real data.
 To obtain a  realistic  estimate of the uncertainties of the cross correlations
we used the bootstrap resampling technique described  by e.g. Ling, Frenk and Barrow 
(1986). 
\section{Results}
\begin{figure}[!t]
\begin{center}
\psfig{file=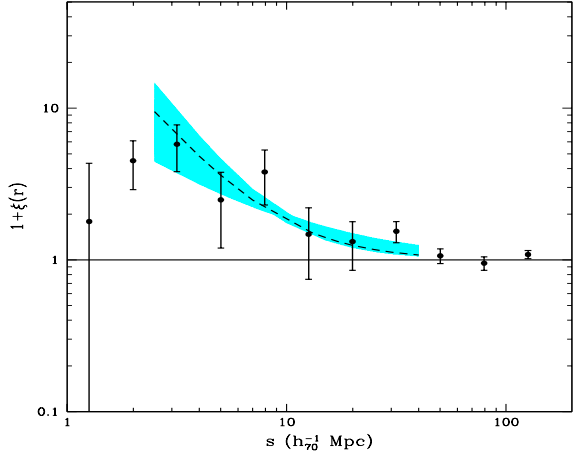,width=8.cm,height=6.5cm,angle=0}
\caption{\label{fig:ccf}The Cluster-AGN soft X-ray cross correlation function plus one. 
The error bars are quoted at 1$\sigma$ level.  The dashed line represents the best fit 
maximum-likelihood power-law fit $s_{0}$=8.7$^{+1.2}_{-0.3}$ $h_{70}^{-1}$ Mpc and 
$\gamma$=1.7$^{+0.2}_{-0.7}$. The shaded region illustrates the 1$\sigma$ confidence 
region of the power-law fit in the distance range in which it was performed.}
\end{center}
\end{figure}
\begin{figure}[!b]
\psfig{file=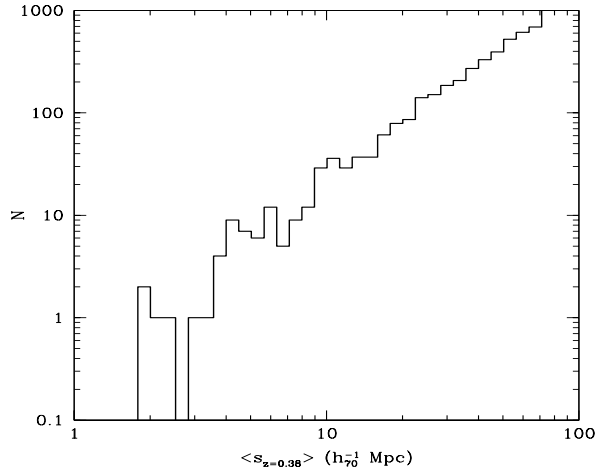,width=8.cm,height=6.5cm,angle=0}
\caption{\label{fig:pairs} The distribution of the angular  radial separations
translated into redshift space distances at $<z>$=0.38 }
\end{figure}
\begin{figure}[!t]
\psfig{file=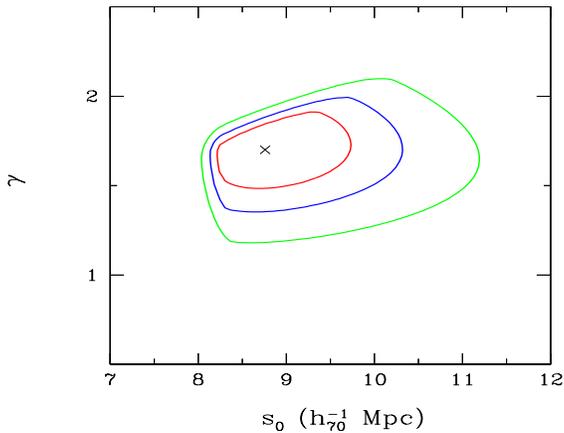,width=8.cm,height=6.5cm,angle=0}
\caption{\label{fig:conts} The 1$\sigma$, 2$\sigma$ and 3$\sigma$ confidence contours 
in the $s_{0},\gamma$ space for the power-law fit to $\xi_{CA}(s)$ for two interesting
parameters. The 
cross  represents the best fit values.}
\end{figure}

We present the spatial cross-correlation function between clusters  and AGNs in 
Fig. \ref{fig:ccf}. A positive clustering signal is detected  in the 
distance interval $s\leq$50 $h_{70}^{-1}$ Mpc. In order to test the strength of 
the clustering we performed a canonical power-law fit,
\begin{equation}
\xi_{CA}(s)=\left(\frac{s}{s_{0}}\right)^{-\gamma},
\label{power}
\end{equation}
with $s_{0}$  and $\gamma$  as  free parameters.
 The fit can be done using the 
coarsely binned data in Fig. \ref{fig:ccf} and minimizing the $\chi^{2}$ 
statistics. However, this approach is extremely sensitive to the size  and 
distribution of the bins. In order to  overcome this problem we  performed a 
standard maximum likelihood power-law fit to unbinned data. 
The comoving separation $s$ was parsed in very small 
bins so that there are either 0 or 1 data pairs for 
any given interval. In this regime Poisson probabilities  are appropriate. 
In order to perform the fit, we need to find  the predicted probability distribution of
the cluster-AGN pairs for each value of $\gamma$ and $s_{0}$. 
We calculated the number of predicted pairs by replacing $\xi_{CA}(s)$ in Eq. \ref{eq:ccf} 
with the model  given by Eq. \ref{power} and  using $D_{C}D_{A}(s)$ (hereinafter $\lambda(s)ds$) 
as variable.
We can then use the separations of all the N cluster-AGN pairs  to  form a likelihood function. 
This is defined as the product of the probabilities of having exactly one pair 
in the interval $ds$ at each separation $s_{i}$ of the N pairs times the probability of having no
pairs in all the other differential intervals.   This for all $s$ in a chosen range ($s_{a}-s_{b}$),
in our case where $\xi_{CA}(s)$ can be reasonably represented by a power law. Assuming Poisson
probabilities we thus obtain the likelihood
\begin{equation}
{\mathcal{L}}=\prod_{i}^{N} \lambda(s_{i})ds\,\exp\sum_{j \neq i}{}\lambda (s_{j})ds,
\end{equation}
where $\lambda(s_{i})ds$ is the expected number of pairs in the interval $ds$, and the index
$j$ runs over all the elements $ds$ which do not contain pairs. We then define the usual 
quantity  $S=-2\ln~{\mathcal{L}}$ and drop the terms independent of model parameters (see e.g. 
\cite{sch98,cro97}) leading to
\begin{equation}
S=2\int_{s_{a}}^{s_{b}}\lambda(s)~ds-2\sum_{i}^{N}\ln[\lambda(s_{i})].
\end{equation} 
 In order to check in which range of separations we can conduct our analysis, 
 we transformed the angular separations between clusters and AGNs into an average redshift space
 separation. We assumed that all the sources were at $<z>$=0.38 (i.e. the median redshift of the 
cluster and AGN sample).
 The result is plotted in Fig. \ref{fig:pairs}. As one can see there are no real pairs
 with separation $<$ 2  $h_{70}^{-1}$ Mpc. For this reason we decided not to consider points 
 at separation lower than 2.5 $h_{70}^{-1}$ Mpc. These  points in the cross-correlation function 
 are mainly introduced
 by the parameters $R_{C}D_{A}$, $R_{A}D_{C}$ in Eq. \ref{eq:ccf} which could have smaller 
 separations than the 
 real data since the random sample   includes neither the extended emission of the clusters nor 
 the broadening 
 due to the PSF of pointlike sources. This prevents us also from overestimates of the correlation length  
 introduced by the amplification bias (see e.g \cite{vik}).
  For this reason  and since on scales larger than 50  $h_{70}^{-1}$ Mpc there is no 
  evidence of signal,  the fit was performed over the distance range  2.5--50 $h_{70}^{-1}$ Mpc.
 In Fig. \ref{fig:conts} we show the  results of the 
maximum-likelihood power-law fit to  $\xi_{CA}(s)$ for the {\em ROSAT} NEP 
survey.  The  1, 2 and 3$\sigma$ were obtained at $\Delta~S$ levels of 2.3, 6.2 and 11.8
 from the minimum value of $S$.    
The best fit parameters obtained are 
$s_{0}$=8.7$^{+1.2}_{-0.3}$ $h_{70}^{-1}$ Mpc and 
$\gamma$=1.7$^{+0.2}_{-0.7}$  where the uncertainty is at the 1$\sigma$
confidence level. 
With $\gamma$ fixed to 1.8 (i.e. a typical value found in galaxy-galaxy correlation function)
 we find 
$s_{0}\sim 8.5 h_{70}^{-1}$ Mpc, a similar value was obtained by extending the fitting region
 to 60 $h_{70}^{-1}$ Mpc and restricting it to the 2.5--40 $h_{70}^{-1}$ Mpc.
 
\noindent
The  integral constraint (\cite{pee80}), which is a systematic shift 
in  correlation functions introduced by the limited volume observed, 
was computed  following the 
prescription of \cite{ro93}. This can be obtained numerically  with a fit  by 
assuming that the correlation function is represented by a power law of fixed 
index (here we used $\gamma$=1.7) on  all scales sampled by the survey. 
The  underestimate of $\xi_{CA}(s)$ due to the integral constraint is 
found to be of order 2\%. \\
In order to determine the stability of these results the procedure was
repeated first by separating the field in two subfields twice, the
North and South, and West and East parts of the survey. The
fluctuations due to sample variance are found to be smaller than the
typical amplitude of the uncertainties.  A similar result is obtained
by recomputing the $\xi_{CA}(s)$ for clusters with $z\leq$0.18
(i.e. the cluster median redshift) and $z>0.18$. The dependence of
$\xi_{CA}(s)$ on the cluster X-ray luminosity ($L_{X}$) was evaluated
dividing the cluster sample into two subsamples with L$_{X}\leq (or
\geq) 3.8\times 10^{43}$~h$_{70}^{-2}$~erg~s$^{-1}$.  
 Though no significant L$_{X}$ dependent behaviour was detected,
we cannot yet conclude that there is no luminosity or redshift
dependent cross-correlation length because of the low-statistics of
the sub-samples. The stability of the result was also checked
fitting the data with the likelihood estimator used by Mullis
et~al.~(2004) returning no significant deviations from our results at
1-2 $\sigma$ level.

\section{Discussion}

We presented here the first direct evidence of spatial clustering of soft X-ray 
selected AGNs around X-ray selected clusters of galaxies. Indirect evidence 
was presented by  Henry~et~al.~(1991), Cappi~et~al.~(2001), 
Cappelluti~et~al.~(2005) (and references therein).  These authors  found 
significant X-ray point source overdensities (about a factor 2) around distant
clusters of galaxies when compared to cluster-free fields. If the 
overdensities were at the cluster redshift they would arise at scales 
smaller than $\sim$7 $h_{70}^{-1}$ Mpc. Since the correlation function is proportional 
to $\left(\frac{\delta\rho}{\rho}\right)^{2}$, a $\xi_{CA}$=1 implies an 
overdensity of a factor 2  with respect to a randomly distributed field. 
We can conclude that,  since the correlation length found in this work  
reflects the scale of the overdensities known up to now, we observe a physical 
overdensity (of at least a factor 2)  of AGNs around clusters between 
2 and $\sim$~8 $h_{70}^{-1}$ Mpc  from the center of the clusters.\\
Because of the shallowness of the NEP survey, the AGN surface density 
(i.e. $<$30 deg$^{2}$ in the central region)
does not allow detection of such a correlation via overdensity analysis since it 
would be dominated by small number statistics. In fact, from our results we expect to detect 
AGNs overdensities on scales $<$7-8 $h_{70}^{-1}$ Mpc from the center of clusters. At $<z>\sim$0.18 (i.e. the 
median $z$ of the cluster sample of the NEP survey) these overdensities arise on scales
of $\sim$0.6 deg$^{-2}$ which are easily resolved by the  NEP survey.  
However to significantly detect these  overdensity on single clusters,
a conspicuous number of sources is necessary
 to disentangle real overdensities 
from  shot noise. 
This problem could be easily resolved
by  high angular resolution telescopes like  
{\em Chandra} and  partially by XMM-{\em Newton}.
  In this direction  
deep
and wide \footnote[1]{The expected fluxes limit for the C-COSMOS and XMM-COSMOS 
survey are 1-2$\times$10$^{-16}$ erg cm$^{-2}$ s$^{-1}$  
and  $\sim$6$\times$10$^{-16}$ erg cm$^{-2}$ s$^{-1}$, respectively. 
These surveys will cover 0.9  deg$^{2}$ and 2.1 deg$^{2}$, respectively
} {\em Chandra} and XMM-{\em Newton} surveys like COSMOS,
which will return an AGN surface density of up to $\sim$ 2700 deg$^{-2}$, 
would allow seeing in a 0.015 deg$^{2}$ 
region (i.e. the size of an ACIS-I chip)  a population of at least 40 AGNs belonging to 
the cluster environment (i.e.   assuming an average overdensity of a factor 2, we 
expect $\sim$40 sources belonging 
to the cluster and $\sim$40 to the background). 
As an example, at the limiting flux of the C-COSMOS survey, the AGN
population of a cluster at $z$=1 would be observed with  a 0.5--2 keV limiting luminosity L$_{min}\sim$10$^{42}$ erg s$^{-1}$.
At $z$=1 the size of an ACIS chip (i.e. 8 arcmin) corresponds to a linear dimension 
of $\sim$4 $h_{70}^{-1}$ Mpc. Having 40 AGNs in a sphere with this  radius, corresponds to a space density of AGN
 with L$_{X}>$L$_{min}$ of 
$\sim$0.15   $h_{70}^{3}$ Mpc$^{-3}$.
We can however state that, according to the result presented here, 
clusters of galaxies could be detected by AGN overdensities (rather than
galaxy overdensities) if the depth of the survey would provide 
an AGN surface density sufficient to overcome the Poisson noise on the AGN number. 
 In general in order to understand the galaxy evolution in dense environments 
the measure of cross correlation between clusters and different kind of galaxies is an
 important tool. We already know that infrared dusty galaxies avoid dense environments 
 therefore showing a large cross-correlation length and weak clustering signal in the small 
 separations region (S\'anchez et al., 2005).
We also know that blue galaxies avoid low-$z$ rich clusters cores (Butcher \& Oemler, 1984). 
It is therefore important to compare the cluster-AGN cross correlation length to that of 
cluster and different galaxy type. 
Mo et al. (1993) computed the cross-correlation function of Abell clusters
and QDOT IRAS galaxies.  They found an average correlation length and
a slope in agreement with the results presented here. Moreover
\cite{lil} showed that the cross-correlation function of Abell
clusters with Lick galaxies is positive on scales $\leq$29
$h_{70}^{-1}$ Mpc with a slope $\gamma\sim$ 2.2 and a correlation
length of $\sim$ 12.6 $h_{70}^{-1}$ Mpc.  These results are also in
agreement within 1 $\sigma$ with our findings on AGNs. These
first comparisons already suggest that AGNs are clustered around
galaxy clusters just like galaxies. 
As a final check we  
compared  our $\xi_{CA}$ to the X-ray cluster-galaxy   cross-correlation 
function (hereinafter CGCCF) computed by \cite{san05}. 
They  used the X-ray selected clusters of the REFLEX  survey (B\"ohringer et~al. 2002)
and the galaxies from the APM survey (\cite{mad90}) limited to b$_{j}$=20.5 mag. 
 They found that the CGCCF behaves like a broken power-law with a cut-off
distance of $\sim$2 $h_{70}^{-1}$ Mpc with a steeper slope at small distances. 
We can define the following approximate biasing relations:
\begin{equation}
\xi_{CA}(s)=b_{C} b_{A} \xi_{\rho}(s),\,\,\,\,\xi_{CG}(s)=b_{C}b_{G}\xi_{\rho}(s).
\end{equation}
Here $\xi_{\rho}(s)$ is the autocorrelation of matter, $b_{G},b_{A}$ and $b_{C}$
are the bias factors relative to  galaxies, AGNs and clusters, respectively.
By dividing the two equations we can then derive $\frac{b_{A}}{b_{G}}(s)$.
In order to perform this operation several effects must be taken in account.

\begin{figure}[!t]
\psfig{file=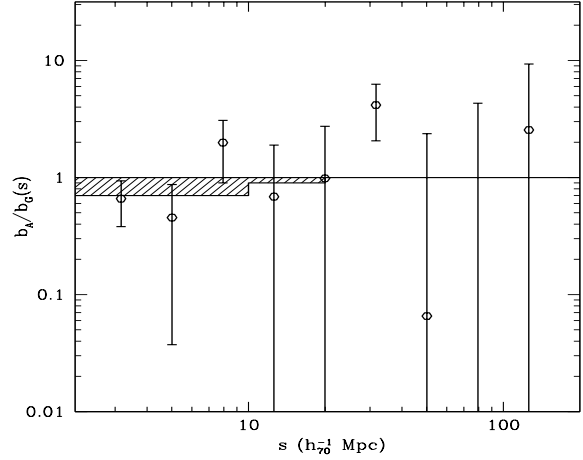,width=8.cm,height=6.5cm,angle=0}
\caption{\label{fig:bias} The ratio between the observed {\em ROSAT} NEP 
 $\xi_{CA}$(s) and the best fit  $\xi_{CG}$(r) obtained by \cite{san05}. 
 Errors  are quoted at the 1$\sigma$ level. The shaded region shows the expected level of 
$\frac{b_{A}}{b_{G}}(s)$=1 if the cross-correlation functions were compared in the same space.}
\end{figure}

\begin{itemize}
\item{The bias of REFLEX clusters could be slightly different from that of 
NEP clusters since they are differently distributed in redshift and the surveys have
different  limiting fluxes and X-ray luminosity distributions. 
\cite{sch01} showed with the power spectrum that even
having a large sample of clusters as REFLEX the error on the bias determination 
is still high. Since in the NEP survey we expect an even higher uncertainty it is a 
reasonable approximation to consider the ratio $\frac{b_{NEP}}{b_{REF}}$ consistent to 1.}
\item{ S\'anchez et al. (2005) computed $\xi_{CG}$ in real space while  
we work in  redshift space. In order to evaluate the effect of redshift space
distortion on  $\xi_{CG}$ we used the results of \cite{cro99}. Their Figs. 3 and 5
 show $\xi_{CG}$ computed both in redshift and real space. From that work we estimate
that our  relative bias  is affected by a $\sim$30$\%$ overestimate below 10 $h_{70}^{-1}$ Mpc
and of  $\sim$10$\%$ between  10 $h_{70}^{-1}$ Mpc and 20 $h_{70}^{-1}$ Mpc. Their work also indicates 
 that $\xi_{CG}$ does
not depend on  the richness of the clusters used in the calculation and that the correction for 
the scale independent biasing on large scales  (\cite{ka}) can be neglected when compared to the size of our uncertainties.} 
 \end{itemize}
\noindent
The ratio $\frac{b_{A}}{b_{G}}(s)$  is plotted in Fig. \ref{fig:bias} as a function of
the distance from the center of the cluster. The shaded region shows the value of our
measurement that implies that 
$\frac{b_{A}}{b_{G}}(s)$=1 when taking into account the difference between real and 
redshift space measurements discussed in the previous paragraph.
 The ratio is consistent with 1 on almost all  scales. We cannot exclude, 
within the errors, much different  values of the relative bias. Our data suggest an 
average relative bias consistent with unity
but allow an upper limit of $\sim$6  (at 1$\sigma$) at separations $s<50$ $h_{70}^{-1}$ Mpc. For separations
$s>10$ $h_{70}^{-1}$ Mpc no lower limits bigger than zero can be given. On 
larger scales the error increase  thus it is  difficult to draw any conclusion. At large separations
the power-law shape of $\xi_{CG}$ becomes uncertain, this  makes a comparison of our data
with those of  S\'anchez et~al.~(2005)  less meaningful.
We cannot exclude a significant antibiasing of AGNs when compared to galaxies, especially 
at low separations.
Though the amplitude of the uncertainties of our data still allows a fluctuation in the relative 
biasing of more than a factor 2, we can  conclude with a precision of 1$\sigma$ that the 
probability for a galaxy to become an  AGN is constant in the range of separations sampled in 
this work and that AGNs can be considered as good tracers of the dark matter distribution 
as are galaxies. 
New deep  and wide field surveys  
such as  XMM-COSMOS (\cite{has06,cap06}) and {\em Chandra}-COSMOS (P.I.: Martin Elvis) with 
their spectroscopically identified  $>$2000 AGNs and $\sim$100 clusters
will improve the precision of this work and will allow us to investigate the 
behaviour of AGNs also in the inner regions of the clusters. In this region the observations 
suggest that the AGNs source density is affected by other physical effects 
such as merging  (\cite{rud05}, Branchesi et al. 2006) or ram pressure phenomena
\section{Summary}
We derived for the first time the soft X-ray spatial cross-correlation
function between clusters and AGNs using the data of the {\em ROSAT}
NEP survey.  A strong clustering signal was detected on scale $s<$50
$h_{70}^{-1}$ Mpc. The best power-law fit parameters are
$s_{0}$=8.7$^{+1.2}_{-0.3}$ $h_{70}^{-1}$ Mpc and
$\gamma$=1.7$^{+0.2}_{-0.7}$.  In this work we observed that the
source density of AGNs is higher near clusters than in the field. This
result confirms earlier findings of overdensities of AGNs around
clusters reported by many authors and improves the evidence
connecting the overdensities to the large scale structure of the
Universe.  We also derived the relative bias between AGNs and galaxies
which is consistent with one on almost all  scales investigated
here. This result still allows, within the errors, a factor 2
fluctuation.  New wide field surveys (such as XMM-COSMOS) performed
with the new generation X-ray telescopes will be useful to enlarge the
statistics, to better understand the physics of AGNs in clusters and
to extend the analysis to the inner regions of clusters.
\begin{acknowledgements}
NC thanks G\"unther Hasinger for the useful discussions, and
Marica Branchesi for an advance communication of her results. 
E.P. is an ADVANCE fellow (NSF grant AST-0340648), also supported by 
NASA grant NAG5-11489. JPH thanks the Alexander von Humboldt Foundation
for support to visit the MPE. We also thank the anonymous referee for her/his 
useful comments and suggestions.\\
{\em In memory of Peter Schuecker.}
\end{acknowledgements}

{}
\end{document}